\documentclass[twocolumn,english,twocolumn,prb,amssymb,amsmath,superscriptaddress,showpacs]{revtex4}
\usepackage[latin9]{inputenc}
\usepackage{graphicx,color}
\usepackage{amssymb}
\usepackage[bookmarks]{hyperref}
\usepackage{babel}

\newcommand{\be}{\begin{equation}}
\newcommand{\ee}{\end{equation}}

\newcommand{\Ls}{_{\scriptscriptstyle{\Lambda}}}
\newcommand{\LdLs}{_{\scriptscriptstyle{\Lambda+\Delta\Lambda}}}
\newcommand{\bis}{^{^{\scriptscriptstyle{>}}}}
\newcommand{\sms}{^{^{\scriptscriptstyle{<}}}}

\newcommand{\knot}{^\mathrm{\scriptscriptstyle{0}}}
\newcommand{\res}{^\mathrm{\scriptscriptstyle{R}}}
\newcommand{\LO}{\!^{^{\Lambda,\Omega}}}

\begin{document}

\title{Nonequilibrium dynamical renormalization group: Dynamical crossover from weak to infinite randomness in the transverse-field Ising chain}

\author{Markus Heyl}
\affiliation{Institute for Quantum Optics and Quantum Information of the Austrian Academy of Sciences, 6020 Innsbruck, Austria}
\affiliation{Institut f\"ur Theoretische Physik, Technische Universit\"at Dresden, 01062 Dresden, Germany}
\affiliation{Institute for Theoretical Physics, University of Innsbruck, 6020 Innsbruck, Austria}
\author{Matthias Vojta}
\affiliation{Institut f\"ur Theoretische Physik, Technische Universit\"at Dresden, 01062 Dresden, Germany}

\begin{abstract}

In this work we formulate the nonequilibrium dynamical renormalization group (ndRG). The ndRG represents a general renormalization-group scheme for the analytical description of the real-time dynamics of complex quantum many-body systems. In particular, the ndRG incorporates time as an additional scale which turns out to be important for the description of the long-time dynamics.  It can be applied to both translational invariant and disordered systems. As a concrete application we study the real-time dynamics after a quench between two quantum critical points of different universality classes. We achieve this by switching on weak disorder in a one-dimensional transverse-field Ising model initially prepared at its clean quantum critical point. By comparing to numerically exact simulations for large systems we show that the ndRG is capable of analytically capturing the full crossover from weak to infinite randomness. We analytically study signatures of localization in both real space and Fock space.

\end{abstract}
\pacs{75.10.Pq,72.15.Rn,05.70.Ln}
\date{\today}
\maketitle


\section{Introduction}

In equilibrium, renormalization group (RG) approaches constitute one of the central concepts for the theoretical description and understanding of many-body systems. One major challenge in the field of nonequilibrium physics~\cite{Polkovnikov2011kx} is the development of appropriate out-of-equilibrium generalizations. Recently, RG techniques have been developed~\cite{Kehrein2006,Berges2008nf,Mitra2012,Mathey2010,Chiocchetta2015} that transfer the idea of scale separation to the out-of-equilibrium dynamics of homogeneous quantum many-body systems. For spin systems in the presence of strong disorder, RG techniques have
been formulated,\cite{Vosk2012,Vosk2013,Pekker} applicable to initial states with
weak entanglement, that extend concepts of the strong-disorder RG~\cite{DasguptaMa,Bhatt1984,Fisher1994} to the nonequilibrium dynamical regime.
Although these RG methods constitute a key step towards the analytical description of the nonequilibrium dynamics in complex systems, controlling the long-time properties is still a major challenge. 

In this work we present a novel nonequilibrium dynamical renormalization group (ndRG) technique for the analytical description of the quantum real-time evolution in complex systems. The ndRG provides a general iterative RG prescription for the full time-evolution operator without the need of diagonalizing the complete Hamiltonian. The ndRG is applicable to both homogeneous as well as disordered quantum many-body systems. Importantly, the ndRG incorporates time as an additional scale which turns out to be important for the description of the long-time dynamics. The main idea behind the ndRG is to separate resonant from off-resonant processes on the basis of the energy--time uncertainty relation:
\be
   \Delta \varepsilon \Delta t \gtrsim \frac{\hbar}{2},
\label{eq:uncertainty}
\ee
that expresses a fundamental limit onto the law of conservation of energy within a
scattering process monitored over a time span $\Delta t$ (Ref.~\onlinecite{Landau1991}). In the asymptotic long-time regime energy-conserving processes dominate the dynamics as apparent, for example, in Boltzmann equations which describe the final asymptotic relaxation to thermal states in homogeneous systems. For long but finite times $t$, however, energy fluctuations are possible such that not only precisely energy-conserving processes contribute but also all those with energy transfer $\Delta \varepsilon \lesssim \hbar/(2t)$ as dictated by the energy--time uncertainty relation.  Therefore, we will classify all processes with $\Delta \varepsilon \lesssim \hbar/(2t) $ as ``resonant'' in the following. The ndRG takes advantage of this limited energy resolution by isolating resonant processes on the basis of a general factorization property of the time-evolution operator. This allows to treat the dynamics of these resonant processes, although nonperturbative in nature, analytically.

We demonstrate the capabilities of the ndRG by applying it to quantum quenches in the disordered transverse-field Ising chain. We study the system's critical dynamics by quenching the system from its clean to its infinite-randomness critical point. We achieve this by preparing the system in the ground state of its homogeneous critical point and then studying its dynamics in presence of weak disorder. We characterize the resulting localization dynamics in both real as well as Fock space. Within the recently developed concept on many-body localization~\cite{Altshuler1997hx,Basko2006,Nandkishore2014,Altman2014}, it is particularly interesting that Fock-space localization has been related to the fundamental question of quantum ergodicity~\cite{Altshuler1997hx} and therefore to thermalization of quantum many-body systems~\cite{Polkovnikov2011kx}. We show that the ndRG is capable of describing the dynamics also in cases when the initially weak perturbation flows to strong coupling.

The remainder of this article is organized as follows. In Sec.~\ref{sec:ndRG} we introduce the ndRG. First, we outline the general idea and present the resulting ndRG equations in Sec.~\ref{sec:ndRGRecipe} that can be applied directly to any model system of interest. The derivation of the ndRG is shown in detail in Sec.~\ref{sec:ndRGDerivation}. In Sec.~\ref{sec:Ising} we then apply the ndRG to the disordered transverse-field Ising chain.


\section{ndRG -- nonequilibrium dynamical renormalization group}
\label{sec:ndRG}

The ndRG, formulated below, is an iterative coarse-graining procedure. It is designed to provide analytical access to the full time-evolution operator of complicated many-body problems without the need of diagonalizing the complete Hamiltonian. As its main goal it isolates resonant from offresonant processes as dictated by the energy--time uncertainty relation, see Eq.~(\ref{eq:uncertainty}). The offresonant processes are eliminated based on scale separation analogous to RG procedures in equilibrium. The resonant processes,  however, that are nonperturbative in nature, cannot be eliminated in this way and, if relevant, can drive the system to strong coupling. Based on a general decoupling mechanism for time-evolution operators we show that the \emph{dynamics} of the resonant processes, although nonperturbative, can still be accessible analytically on all time scales. This is possible even though the system might flow to strong coupling as we will demonstrate for the disordered transverse-field Ising chain in Sec.~\ref{sec:Ising}.

\subsection{The ndRG recipe}
\label{sec:ndRGRecipe}

Consider a system whose Hamiltonian
\be
	H\Ls=H\knot\Ls +V\Ls,
\ee
at a given ultraviolet (UV) cutoff $\Lambda$ can be decomposed into an exactly solvable part $H\knot\Ls$ and a weak perturbation $V\Ls$, with an associated time-evolution operator
\be
	P\Ls(t) = \exp[-iH\Ls t].
\ee
Notice
that $\Lambda$ could either be chosen as a momentum or energy cutoff.
Given that the UV cutoff takes a value $\Lambda+\Delta \Lambda$ we now aim at reducing it to $\Lambda$
by eliminating contributions $V\bis\LdLs$  of $V\LdLs=V\bis\LdLs +
V\sms\LdLs$ involving particles with momenta $q$ in the shell $|q|\in
[\Lambda,\Lambda+\Delta \Lambda]$ and thereby generating a renormalized effective theory for the remaining modes. Operators with a subscript  $\Lambda$, e.g., $H\knot\Ls$, denote the renormalized ones after the RG step from $\Lambda+\Delta \Lambda$ to $\Lambda$ has been made while those with a subscript $\Lambda +\Delta \Lambda$ refer to the initial operators before the RG step is performed.

Importantly, we do not aim at integrating out all contributions of modes in this shell, but only those which are offresonant according to Eq.~(\ref{eq:uncertainty}). The resonant processes remaining in the Hamiltonian are finally dealt with on the basis of a general factorization property of time-evolution operators. This is the main feature of the ndRG and distinguishes it from other RG schemes.
The ndRG eliminates the offresonant contributions from the time-evolution operator $P\LdLs(t) = \exp[-i H\LdLs t]$ by constructing explicitly a unitary transformation yielding
\be
	P\LdLs(t) = e^{-S\Ls}P\Ls(t) e^{S\Ls}.
\ee
with $P\Ls(t) = \exp[-i H \Ls t]$ containing the dynamics of the remaining degrees of freedom.
For a given Hamiltonian, the ndRG scheme works according to the following prescription which is straightforward to implement for a given model system:
\begin{enumerate}
 \item Identify the off-resonant processes $V\LdLs\bis$ and decompose:
	\be
		V\LdLs=V\bis\LdLs + V\sms\LdLs.
	\ee
 \item Determine the generator $S\Ls$ of the unitary transformation via
	\be
		S\Ls-S\Ls(t) = -i \int_0^t dt' V\bis\LdLs (t').
		\label{eq:generatorDef}
	\ee
	Here, $V\bis\LdLs(t) = P\Ls^{\mathrm{\scriptscriptstyle{0}} \dag}(t) V\bis\LdLs P\knot\Ls(t) $ with $P\knot\Ls(t)=\exp[-iH\knot\Ls t]$ the free time-evolution operator.
 \item Obtain the renormalized Hamiltonian:
	\be
		H\Ls = H\knot\LdLs + V\sms\LdLs + \frac{1}{2} [S\Ls,V\bis\LdLs].
		\label{eq:recipeHamRen}
	\ee
 \item Iterating the above steps by successively lowering the cutoff $\Lambda$ one obtains at the end of the ndRG transformation the following representation of the full time-evolution operator
	\be
		P(t) = U_\ast^\dag e^{-iH_\ast t} U_\ast, \quad H_\ast = H_\ast\knot + V_\ast^\mathrm{R},
	\ee
 with $H\knot_\ast$ denoting the renormalized exactly solvable part and $V_\ast^\mathrm{R}$ the remaining resonant contributions. The unitary transformation $U_\ast$ is a $\Lambda$-ordered exponential:
	\be
		U_\ast = \mathcal{T}_\Lambda e^{ \int_{\Lambda_\ast}^{\Lambda\knot} S_\Lambda},
		\label{eq:Uast}
	\ee
  with $\Lambda\knot$ the initial and $\Lambda_\ast$ the final UV cutoff. $\mathcal{T}_\Lambda$ denotes $\Lambda$-ordering analogous to common time ordering.  The \emph{dynamics} of this apparently complicated problem can  be solved on the basis of a factorization property of the time-evolution operator for resonant processes:
 \be
	e^{-i H_\ast t}  \approx e^{- i H_\ast\knot t} e^{-i V_\ast^\mathrm{R} t}.
	\label{eq:recipeTEOF}
 \ee
\end{enumerate}
These equations are valid up to second order in the perturbation strength, the extension to higher orders is straightforward.
In what follows, we will present the derivation of the ndRG procedure. For readers interested in the application of the ndRG for a concrete model system, it is possible to directly consult Sec.~\ref{sec:Ising} where the ndRG is applied to the disordered transverse-field Ising model.

\subsection{Derivation of the ndRG}
\label{sec:ndRGDerivation}

As anticipated before, consider a Hamiltonian $H\LdLs = H\LdLs\knot + V\LdLs$ that can be separated into an exactly solvable part $H\LdLs\knot$ and a weak perturbation $V\LdLs$ at a UV cutoff $\Lambda+\Delta \Lambda$.
In order to derive the ndRG procedure, let us first turn to an interaction picture with respect to the free Hamiltonian $H\knot\Ls$ at the desired final cutoff $\Lambda$ after the RG step has been performed. Of course, $H\knot\Ls$ is not known a priori but has to be determined self-consistently in the end which is straightforward to implement. In the interaction picture the time-evolution operator is:
\be
	P_{\scriptscriptstyle{\Lambda+\Delta \Lambda}}(t) = e^{-i H\knot\Ls t} \,\, W\Ls(t),
\label{eq:interaction_picture}
\ee
where
\be
	W\Ls(t) = T \exp\left[ -i \int_0^t dt'  \Big[ H_{\scriptscriptstyle{\Lambda+\Delta \Lambda}} (t')  -H\knot\Ls (t') \Big]   \right],
\ee
with $H_{\scriptscriptstyle{\Lambda+\Delta \Lambda}}(t) = \exp[iH\knot\Ls t] H_{\scriptscriptstyle{\Lambda+\Delta \Lambda}} \exp[-iH\knot\Ls t] $ and $T$ denotes the usual time-ordering prescription. In the following, time-dependent operators $\mathcal{O}(t)$ will always refer to being time evolved in the interaction picture, i.e., $\mathcal{O}(t) = \exp[iH\knot\Ls t]\mathcal{O} \exp[-iH\knot\Ls t] $.

\subsubsection{Disentangling theorem}

The interaction picture representation of the time-evolution operator $P_{\scriptscriptstyle{\Lambda+\Delta \Lambda}}(t)$ in Eq.~(\ref{eq:interaction_picture}) is still exact, for the complicated models of interest, however, the time-ordered exponential in $W\Ls(t)$ cannot be evaluated easily. The goal of the RG procedure is not to find an approximate solution to $W\Ls(t)$ as a whole. Instead, we aim at an iterative sequence of transformations by successively integrating out high-energy degrees of freedom. For that purpose we use a disentangling theorem for time-ordered exponentials:\cite{Kampen1974}
\be
	W\Ls(t) = W\Ls\bis(t) W\Ls\sms(t),
\label{eq:disentangling}
\ee
where
\begin{align}
	W\Ls\bis(t) & = T \exp\left[ -i \int_0^t dt' \mathcal{K}\Ls(t') \right], \nonumber \\
	W\Ls\sms(t) & = T \exp\left[ -i \int_0^t dt' \,\, \mathcal{V}\Ls(t') \right],
\label{eq:Wbig_Wsmall}
\end{align}
and
\begin{align}
	 \mathcal{V}\Ls(t) & = [ W\Ls\bis(t)]^\dag \,\,\Big[ H_{\scriptscriptstyle{\Lambda+\Delta \Lambda}}(t)-H\knot\Ls(t) -\mathcal{K}\Ls(t) \Big] \,\, W\Ls\bis(t).
\label{eq:mathcalV}
\end{align}
While the operator $W\Ls\bis(t)$ will be chosen such to eliminate the desired processes, the operator $\mathcal{V}\Ls(t)$ will determine the renormalized Hamiltonian after the RG step. The precise choice of the operator $\mathcal{K}\Ls$ for the purpose of the ndRG will be given below.

The aim of the ndRG is to find a representation of $W\bis\Ls(t)$ of the following form:
\be
	W\Ls\bis(t) = e^{-S\Ls(t)}e^{S\Ls},
\ee
for some suitable antihermitian operator $S\Ls$ such that $\exp[S\Ls]$ is unitary. Before going into the details of how to derive this identity we first would like to illustrate its main consequences.
Using this identity, we obtain
\be
	W\Ls\sms(t)=T\exp\left[-i \int_0^t dt' \,\, e^{-S\Ls} e^{S\Ls(t')}
	\,\,\mathcal{ V}\Ls(t') \,\,e^{-S\Ls(t')} e^{S\Ls}\right],
\ee
which gives, because $S\Ls
$ is independent of time:
\begin{align}
	W\Ls\sms(t) = e^{-S\Ls}  T\exp\left[-i\int_0^t dt' \,\, V\Ls(t')  \right] e^{S\Ls},
\label{eq:W_smaller_end}
\end{align}
by defining
\be
	V\Ls=e^{S\Ls} \Big[ H_{\scriptscriptstyle{\Lambda+\Delta \Lambda}}-H\knot\Ls -\mathcal{K}\Ls \Big]e^{-S\Ls}.
\label{eq:new_perturbation_fo}
\ee
Bearing in mind that, using Eqs.~(\ref{eq:interaction_picture},\ref{eq:disentangling}), the time-evolution operator $P_{\scriptscriptstyle{\Lambda+\Delta \Lambda}}(t)$ has been factorized according to:
\be
	P_{\scriptscriptstyle{\Lambda+\Delta \Lambda}}(t) = e^{-i H\knot\Ls t} W\Ls\bis(t) W\Ls\sms(t),
\ee
one obtains
\be
	P_{\scriptscriptstyle{\Lambda+\Delta \Lambda}}(t) = e^{-iH\knot \Ls t}\,\, e^{-S\Ls(t)}\,\, T\exp\left[-i\int_0^t dt' \,\, V\Ls(t')  \right] e^{S\Ls}.
\ee
which then yields
\be
	P_{\scriptscriptstyle{\Lambda+\Delta \Lambda}}(t) = e^{-S\Ls}  e^{-iH\knot \Ls t} \,\, T\exp\left[-i\int_0^t dt' V\Ls(t')\right]e^{S\Ls}.
\ee
Now one can switch back from the interaction to the original picture such that:
\be
	P_{\scriptscriptstyle{\Lambda+\Delta \Lambda}}(t)= e^{-S\Ls} e^{-i [H\knot\Ls + V\Ls]t }e^{S\Ls},
\label{eq:Pfinal}
\ee
which is the desired identity provided $S\Ls$ is chosen such that $H\Ls\knot+V\Ls$ becomes the renormalized Hamiltonian after eliminating the contribution $V\LdLs\bis$, see Eq.~(\ref{eq:recipeHamRen}). In the following, we now show how this can be achieved.

\subsubsection{Magnus expansion}

The crucial point is that $W\Ls\bis(t)$ in Eq.~(\ref{eq:Wbig_Wsmall}) can be evaluated approximately within a controlled expansion. The main complications in computing $W\Ls\bis(t)$ arise from the time ordering prescription that makes it difficult, and for interesting problems impossible, to evaluate exactly. Most importantly, the operator $\mathcal{K}\Ls$  turns out to proportional to the strength of the weak perturbation  such that a Magnus expansion is applicable:
\begin{align}
	W\Ls\bis(t) &= \exp \left[  -i\int_0^t dt' \mathcal{K}\Ls(t') - \right. \nonumber \\ & \left. -\frac{1}{2}\int_0^t dt' \int_0^{t'}dt'' \Big[\mathcal{K}\Ls(t'),\mathcal{K}\Ls(t'')\Big] + \dots \right].
\label{eq:Magnus}
\end{align}
which transforms the time-ordered exponential into a conventional exponential on the expense of an infinite series. Importantly, this expansion is controlled by a small parameter which is the perturbation strength. Here, the Magnus expansion is shown up to second order which is sufficient for the targeted accuracy. In case of a larger desired precision, higher orders of the Magnus expansion have to be included.

\subsubsection{Generator of the ndRG transformation}

In order to transform $W\bis\Ls(t)$ into the desired form we choose:
\begin{align}
	\mathcal{K}\Ls = V\bis\LdLs - \frac{1}{2} \left[ S\Ls,V\bis\LdLs \right]
	\label{eq:choice_K}
\end{align}
with $S\Ls$ given by the simple integral
\be
	S\Ls - S\Ls(t) = -i \int_0^t dt' \,\, V\bis\LdLs(t').
\ee
Inserting this into the Magnus expansion, see Eq.~(\ref{eq:Magnus}), one obtains taking into account all contributions up to second order in the perturbation strength
\be
	W\Ls\bis(t) = e^{S\Ls-S\Ls(t)+\frac{1}{2}[S\Ls,S\Ls(t)]},
\ee
which, using the Baker-Campell-Hausdorff formula, is equivalent to the desired expression:
\be
	W\Ls\bis(t) = e^{-S\Ls(t)}e^{S\Ls},
\ee
again taking into account all contributions up to second order in the perturbation strength.

\subsubsection{Renormalized Hamiltonian}

Having established the generator $S\Ls$ of the ndRG, it remains now to determine the renormalized Hamiltonian after the RG step. For that purpose we use that the generator $S\Ls$ of the transformation is proportional to the perturbation strength such that the following expansion for Eq.~(\ref{eq:new_perturbation_fo}) is applicable:
\be
	V\Ls = H_{\scriptscriptstyle{\Lambda+\Delta \Lambda}}-H\knot\Ls -\mathcal{K}\Ls + \Big[ S\Ls,H_{\scriptscriptstyle{\Lambda+\Delta \Lambda}}-H\knot\Ls -\mathcal{K}\Ls \Big] + \dots
\ee
Using the choice for $\mathcal{K}\Ls$ in Eq.~(\ref{eq:choice_K}) this  gives
\begin{align}
	V\Ls = & H\LdLs\knot - H\Ls\knot + V\LdLs\sms + \nonumber \\ & + \frac{1}{2} \left[ S\Ls,V\bis\LdLs \right] + \left[ S\Ls, V\LdLs\sms \right],
\end{align}
taking into account all terms up to second order accuracy. According to Eq.~(\ref{eq:Pfinal}), the renormalized Hamiltonian is then given by:
\begin{align}
	H\Ls =& H\Ls\knot + V\Ls= H\LdLs\knot + V\LdLs\sms + \nonumber \\ & + \frac{1}{2} \left[ S\Ls,V\bis\LdLs \right] + \left[ S\Ls, V\LdLs\sms \right].
\end{align}
Importantly, however, the last contribution $\left[ S\Ls, V\LdLs\sms \right]$ can be neglected. This operator contains again contributions of modes in the shell $[\Lambda,\Lambda+\Delta \Lambda]$ as it was for $V\LdLs\bis$, the strength of  $\left[ S\Ls, V\LdLs\sms \right]$, however, is now of second order. Eliminating this contribution in the same way as $V\LdLs\bis$, will then only contribute beyond second order in the perturbation strength in the renormalized Hamiltonian and is therefore beyond the desired accuracy. Concluding, the final renormalized Hamiltonian taking into account all contributions up to second order in the perturbation strength reads:
\be
	H\Ls = H\LdLs\knot + V\LdLs\sms +  \frac{1}{2} \left[ S\Ls,V\bis\LdLs \right],
\ee
which is the desired result presented already in Eq.~(\ref{eq:recipeHamRen}).

\subsection{Factorization of the time-evolution operator}

At the end of the ndRG procedure, one ends up with a Hamiltonian
\be
	H_\ast = H\knot_\ast + V_\ast\res.
\ee
By construction we have not eliminated all processes of the perturbation
$V$, but kept the resonant contributions $V\res_\ast$ that still have to be accounted for.
This seemingly complicated problem, however, can be simplified substantially
\emph{because} the processes in $V\res_\ast$ are resonant as the renormalized time-evolution operator approximately factorizes:
\begin{equation}
   e^{-i H_\ast t/\hbar} \approx e^{-i H\knot_\ast t/\hbar} e^{-i V\res_\ast t/\hbar}.
\label{eq:teo_factorization}
\end{equation}
This property can be seen in the following way. In the interaction picture the renormalized time-evolution operator obeys $\exp[-iH_\ast t/\hbar]= \exp[-iH\knot_\ast t/\hbar] T\exp[-i \int_0^t dt' V\res_\ast(t')/\hbar] $. As $V\res_\ast$ only contains  resonant processes with $\Delta \varepsilon < \hbar/(2t)$ we have that $V\res_\ast(t) = \exp[iH\knot_\ast t/\hbar]V\res_\ast \exp[-iH\knot_\ast t/\hbar] \approx V\res_\ast$ is approximately constant in time leading directly to the factorization in Eq.~(\ref{eq:teo_factorization}).

In typical problems the complexity of $H$ originates from the noncommutativity of
$H\knot$ and $V$ while their individual properties are much easier to determine. The
major advantage of Eq.~(\ref{eq:teo_factorization}) is the separation of the resonant
processes of the perturbation from the dynamics of the (renormalized) unperturbed system
whose individual time evolution can be determined much easier as will be demonstrated for
the Ising model with disorder below.

\subsection{Discussion}

Summarizing, in this section we have introduced the ndRG. In Sec.~\ref{sec:ndRGRecipe} we have presented the ndRG recipe that can be straightforwardly implemented for any model system of interest. In Sec.~\ref{sec:ndRGDerivation} a detailed derivation of the resulting ndRG equations has been given.

Up to now we have not specified when to stop the ndRG transformation. Of course, the ndRG always stops when there are no remaining off-resonant modes as it will occur for the random transverse-field Ising chain, see Sec.~\ref{sec:Ising}. Importantly, we can take advantage of the additional scale time appearing in the ndRG. Specifically, consider a mode of energy $\varepsilon$. As the time-evolution operator only contains the product $\varepsilon t/\hbar$ we have that for times $t<\varepsilon/\hbar$ this mode is essentially inert. In other words, all modes with energies $\varepsilon<\hbar/t$ are frozen out and can be dealt with on a purely perturbative basis using time-dependent perturbation theory. Therefore, we can stop the ndRG transformation when the energy $\varepsilon_\Lambda$ of the modes at the UV cutoff $\Lambda$ reaches $\varepsilon_\Lambda = \hbar/t$ and we obtain a time-dependent final UV cutoff $\Lambda_\ast = \Lambda_\ast(t)$.

This observation has important consequences. In particular, consider a system with a strong-coupling divergence where we leave the region of validity of the ndRG when reducing the UV cutoff too much. Utilizing that for not too large times $t$ the ndRG transformation can be stopped at a UV cutoff $\Lambda_\ast(t)$ which is large enough such that the strong-coupling divergence is not yet effective it is still possible to describe the dynamics of the system. In other words, the dynamics on not too long times is still accessible on the basis of the ndRG although the system might flow to strong coupling in the asymptotic long-time limit.


We would like to emphasize that the ndRG can be generalized also to other temporal dependencies of the Hamiltonian beyond the quantum quench considered here. This is possible because time itself constitutes an essential element of this RG by construction as it is utilized explicitly for the resonance condition, for example. In this context, it might be particularly interesting to study crossovers to the adiabatic limit by considering ramps instead of quenches where universality such as Kibble-Zurek scaling~\cite{Kibble76,Zurek1985,Polkovnikov2011kx} can be observed. Moreover, the ndRG also inherits the potential to extend RG ideas to periodically driven systems. This is of particular interest in view of the recently discovered energy-localization transitions~\cite{DAlessio2012} which represent a novel class of nonequilibrium phase transitions in complex quantum many-body systems.


\section{Quantum quenches in the disordered transverse-field Ising chain}
\label{sec:Ising}

In the previous section we have introduced the ndRG. It is the aim of the following analysis to apply the ndRG to a paradigmatic model system, the one-dimensional transverse-field Ising chain.\cite{Sachdev2011} The transverse-field Ising chain can be solved exactly also in the presence of disorder. Therefore, it is ideally suited to demonstrate the capabilities of the ndRG by comparing to exact numerical simulations. In particular, we aim at demonstrating that the ndRG is capable to describe both the dynamics on intermediate time scales as well as in the long-time limit which is a challenging task.

In noninteracting one-dimensional quantum systems the effect of arbitrarily weak
randomness is substantial: all single-particle eigenstates become
localized.\cite{Abrahams1979} As a consequence, particles separated over distances larger than the localization length cannot exchange information and are therefore essentially unentangled. The localization dynamics for unentangled initial states in one-dimensional systems show universal behavior~\cite{Bardarson2012,Serbyn2013} that can be attributed to a dynamical renormalization-group fixed point~\cite{Vosk2012,Vosk2013} where localization not only happens in real but also in many-body Hilbert space.\cite{Altshuler1997hx,Basko2006} But how is information propagating within a disordered landscape when the system is highly entangled initially? In the remainder of this article we aim at studying this question exemplarily for the disordered transverse-field Ising chain.

\subsection{Model and setup}

To study the localization dynamics out of quantum correlated states we consider a one-dimensional Ising model with transverse-field disorder:
\be
   H = -\frac{J}{2}\sum_{l=1}^N \left[ \sigma_l^z\sigma_{l+1}^z + h_l \sigma_l^x  \right].
\label{eq:RTFIM}
\ee
In equilibrium, the homogeneous system with $h_l=h$ shows a
quantum phase transition at $h=1$ separating a ferromagnetic ($h<1$) from a paramagnetic
($h>1$) phase.\cite{Sachdev2011} According to the Harris criterion~\cite{Harris1974} the
quantum critical point is unstable against weak disorder, and it has been shown that
the system flows to an infinite-randomness fixed point instead.\cite{Fisher1994}

In this paper, we develop a dynamical theory for this flow from weak to infinite
randomness in nonequilibrium real-time evolution where progressing time itself drives
this crossover. The quantum correlated state is initialized by preparing the system in
the ground state $|\psi_0\rangle$ of the clean critical model at $h_l=h=1$. The localization
dynamics is generated by switching on weak disorder suddenly inducing nonequilibrium
real-time evolution that is formally solved by
\be
   |\psi_0(t)\rangle = P(t) \,|\psi_0\rangle,\,\,\, P(t)=e^{-i H t/\hbar}.
\label{eq:propagator}
\ee
The distribution for the random fields $h_l$ is chosen such that $\langle \log(h_l) \rangle_\mathrm{dis}=0$ with $\langle \dots \rangle_\mathrm{dis}$ the disorder average. Thus the ground state of the system is located right at the infinite-randomness critical point.\cite{Fisher1994,Pfeuty1979}

Contrary to typical condensed-matter systems where disorder is
ubiquitous, systems of cold atoms in optical lattices are clean, and disorder has to be
imposed, for example, by laser speckle patterns~\cite{Bloch2008lv} providing an ideal
candidate for the implementation of the anticipated nonequilibrium protocol. Moreover,
the model in Eq.~(\ref{eq:RTFIM}) can also be simulated within circuit
QED~\cite{Viehmann2012} where disorder is also tunable.\cite{Viehmann2013}

Numerically, this model can be solved exactly for large systems.\cite{Lieb1961go} In this work we will present exact results for systems up to $N=3200$ lattice sites. Due to the broken translational invariance in presence of disorder, however, this model is challenging for analytical methods. In equilibrium, this system has been solved exactly in the vicinity of its critical point both in the weak disorder limit~\cite{McKenzie1996} and in the vicinity of the infinite-randomness critical point.\cite{Fisher1994} Out of equilibrium, the dynamics in the vicinity of the infinite-randomness critical point for weakly entangled initial states has been studied analytically recently.\cite{Vosk2013} In the present work, we will address a complementary viewpoint -- the localization dynamics for strongly entangled initial states at weak disorder. In particular, applying the ndRG to this model system will serve as a benchmark for gauging the capabilities of the methodology introduced in this work.

The transverse-field Ising model in Eq.~(\ref{eq:RTFIM}) can be
diagonalized exactly by a mapping to a free fermionic theory using a Jordan-Wigner transformation:\cite{Lieb1961go}
\be
	H = J \sum_l h_l c_l^\dag c_l - J\sum_l \left[ c_l^\dag c_{l+1} + c_l^\dag c_{l+1}^\dag + \mathrm{H.c.} \right].
\ee
with $c_l$ fermionic annihilation operators at site $l=1,\dots,L$. For the numerical implementation we parametrize
\be
	h_l=e^{\eta_l},
\ee
with $\eta_l \in [-\delta,\delta]$ drawn from uncorrelated uniform distributions, i.e., $\langle \eta_l \eta_{l'} \rangle_\mathrm{dis} = \delta_{ll'} \delta^2/3$, yielding $\langle \log(h_l) \rangle_\mathrm{dis}=0$. As emphasized before, this ensures that the system is located right at the infinite-randomness critical point.\cite{Pfeuty1979,Fisher1994} In the analytical treatment we use the parametrization
\be
	g=\langle h_l\rangle_\mathrm{dis},\quad g_l = h_l-\langle h_l \rangle_\mathrm{dis}.
\ee
The strength of the disorder we characterize via the variance $\sigma^2 = \langle g_l^2 \rangle_\mathrm{dis}$ which in the weak-disorder limit becomes $\sigma^2 \to \delta^2/3$. It is important to note that  this parametrization, which is a consequence of the condition $\langle \log(h_l) \rangle_\mathrm{dis}=0$, yields
\be
	g = 1 + \frac{\delta^2}{6}, \quad g_l \in [-\delta,\delta].
	\label{eq:field_analytical_parametrization}
\ee
Therefore, a larger homogeneous critical transverse field $g_c = 1+\sigma^2/2$ is required to destroy the ferromagnetic order as opposed to the case without disorder where $g_c=1$. This is a consequence of the fluctuating local fields which can locally decrease the homogeneous field when $g_l<0$ favoring magnetic order.


\subsection{Summary of results}
\label{sec:resultsSummary}

Based on the ndRG and corroborated by extensive numerical simulations we study the dynamics in the disordered transverse-field Ising chain induced by a quantum quench from the homogeneous to the infinite-randomness critical point. We investigate the resulting localization dynamics from two perspectives, namely through localization in Fock as well as real space. Here, we summarize our main findings, whose derivation and more detailed analysis can be found in Sec.~\ref{sec:results}.


Unlike in classical systems a general understanding of ergodicity for quantum many-body systems has not yet been achieved.\cite{Polkovnikov2011kx} Recently, however, a framework addressing this fundamental problem has been proposed, many-body localization,\cite{Altshuler1997hx,Basko2006,Nandkishore2014,Altman2014} where the transition from ergodic to nonergodic is associated with an Anderson localization transition in Fock space.\cite{Altshuler1997hx} In this work, we quantify the localization in Fock space by studying the temporal deviation of the system from its initial Fock state $|\psi_0\rangle$ via the Loschmidt echo
\be
	\mathcal{L}(t) = \left| \langle \psi_0 | e^{-iHt} | \psi_0 \rangle \right|^2.
	\label{eq:defLoschmidtEcho}
\ee
Due to its large-deviation scaling $\mathcal{L}(t) = \exp[-N\lambda(t)]$~\cite{Silva2008gj,Gambassi2012a,Heyl2013a} with $\lambda(t)$ an intensive function we will consider $\lambda(t)$ in the following. We find that on intermediate time scales the rate function $\lambda(t)$ shows a very slow temporal logarithmic growth:
\be
	\overline{\lambda}(t) = \langle \lambda(t) \rangle_\mathrm{dis}  = \frac{\sigma^2}{\pi^2} \log(2Jt/\hbar),
\ee
for  $\hbar/(J\sigma) \ll t \ll \hbar/(J \sigma^2)$ with $\langle \dots\rangle_\mathrm{dis}$ denoting the average over disorder realizations. An increasing $\overline{\lambda}(t)$ implies an increasing deviation of the time-evolved state from the initial Fock state. The slow logarithmic growth of the Loschmidt rate function we interpret as an indicator of Fock space localization: the system only very slowly departs from its initial condition.

In the asymptotic long-time limit, the rate function approaches a constant value
\be
	 \lambda_\infty  = \lim_{t\to \infty} \overline{\lambda} (t) = \frac{\sigma^2}{\pi^2} \log\left( \frac{2J}{\sigma^2\hbar}\right).
\ee
We find that $\lambda_\infty$ is related to the Fidelity $F=|\langle \psi_0 | \phi_0 \rangle | = \exp[-N f]$ with $|\phi_0\rangle$ the ground state of the final Hamiltonian:
\be
	\lambda_\infty = 4 \langle f \rangle_\mathrm{dis},
\ee
by calculating $F$ numerically exactly for large systems. Therefore, the asymptotic long-time behavior of the Loschmidt echo, containing, in principle, information about the full many-body spectrum, is only determined by equilibrium ground-state properties which we attribute to localization in Fock space.

Importantly, Fock-space localization and
localization in real space are typically connected.\cite{Altshuler1997hx,Basko2006} While in homogeneous systems local
correlations decay in time this is not the case in localized systems.\cite{Anderson1958wx} Signatures for retaining local memory are contained in the long-time behavior of the autocorrelation function~\cite{Anderson1958wx,Iyer2012}
\be
   \chi(t) = \left\langle \frac{1}{N} \sum_{l=1}^N \langle \sigma_l^x(t) \sigma_l^x \rangle_c \right\rangle_\mathrm{dis}
\ee
Here, $\langle  \sigma_l^x(t) \sigma_l^x  \rangle_c = \langle \sigma_l^x(t) \sigma_l^x  \rangle-\langle  \sigma_l^x(t) \rangle\langle \sigma_l^x  \rangle$ denotes the cumulant and $\langle \dots \rangle=\langle \psi_0|\dots |\psi_0\rangle$ the average with respect to the initial state.

Using the ndRG, we show below that the decay of the local memory on intermediate time scales is algebraic:
\be
   \chi(t) = \frac{1}{2} e^{-2iJt/\hbar} \left[ \frac{e^{-i\pi/4}}{(\pi Jt/\hbar)^{3/2}}  - \frac{\sigma^2}{2\pi} \right]
   \label{eq:result_chi}
\ee
for  $\hbar/(J\sigma) \ll t \ll \hbar/(J \sigma^2)$.  Remarkably, the presence of disorder induces a static time-independent contribution (up to a phase factor) signaling a nondecaying local memory  with a weight $\sigma^2$ given by the strength of the random potential. Therefore, the system shows localization complementing the observed Fock-space localization in terms of the Loschmidt echo above for real space.  We confirm the preservation of local memory using numerically exact simulations of the dynamics for the asymptotic long-time limit where we find that that $\chi_\infty = \lim_{t\to\infty} \langle \chi(t) \rangle_\mathrm{dis} \not=0$.

\subsection{ndRG for the disorder transverse-field Ising chain}

Having summarized the main results obtained in this work it is now the aim to show in detail their derivation on the basis of the ndRG. First, we outline the exact analytical diagonalization of the homogeneous system and then we represent the weak disorder perturbation in this basis. In Sec.~\ref{subsec:generator} we construct the generator $S\Ls$ of the ndRG and summarize the resulting ndRG scaling equations for the spectrum and the disorder strength. In Secs.~\ref{sec:derivationSpectrum} and \ref{sec:derivationDisorder} we show how these scaling equations are determined within the ndRG. A detailed treatment of the resonant process will be given in Sec.~\ref{sec:ndRGResonantProcesses}. The derivations of the results for the observables, already summarized above, will be presented in Sec.~\ref{sec:results}.

According to the ndRG recipe given in Sec.~\ref{sec:ndRGRecipe} we first decompose the Ising  Hamiltonian in Eq.~(\ref{eq:Hamiltonian}) into an exactly solvable part $H\knot$ and a perturbation $V$ which in the weak-disorder limit is:
\be
	V = \sum_l g_l c_l^\dag c_l,\quad H\knot = H-V.
\ee
The homogeneous part $H\knot$ can be diagonalized explicitly using  Fourier transformation and a subsequent Bogoliubov rotation:~\cite{Sachdev2011}
\be
	c_k = \frac{1}{\sqrt{N}} \sum_l e^{ikl} c_l = \cos(\theta_k) \gamma_k - i \sin(\theta_k)\gamma_{-k}^\dag,
\label{eq:c_k}
\ee
with $\tan(2\theta_k) =\sin(k)/[g-\cos(k)]$. Then, the Hamiltonian $H\knot$ becomes
\be
	H\knot = \sum_k \varepsilon_k \gamma_k^\dag \gamma_k, \quad \varepsilon_k = J \sqrt{(g-\cos k )^2+\sin^2 k}.
	\label{eq:diagonal_H0}
\ee
In the present case where $g=1+\sigma^2/2$, see Eq.~(\ref{eq:field_analytical_parametrization}), the Bogoliubov angles $\theta_k$ can be determined asymptotically which yields
\be
	\theta_k = \left\{ \begin{array}{ll}
	                    (\pi-k)/4 & , \mathrm{for} \,\,\, k\gg \sigma^2 \\
	                    ~&~\\
	                    k/(2\sigma^2) & ,\mathrm{for} \,\,\, 0 \leq k \ll \sigma^2
	                   \end{array}
	\right.
	\label{eq:theta_k}
\ee
with $\theta_{-k} = -\theta_k$. In this  basis the disorder contribution $V$ to the Hamiltonian reads
\begin{align}
   V  =& \sum_{kk'} \omega_{kk'} \gamma_k^\dag \gamma_{k'} + \sum_{kk'} \left[ m_{kk'} \gamma_k^\dag \gamma_{k'}^\dag + \mathrm{h.c.}  \right],
\label{eq:Hamiltonian}
\end{align}
where
\begin{align}
	\omega_{kk'} & = Jg_{k-k'}\cos[\theta_k+\theta_{k'}], \nonumber \\ m_{kk'} & =-iJg_{k+k'}\sin[\theta_k-\theta_{k'}]/2,
	\label{eq:wmAmplitudes}
\end{align}
with $g_k = N^{-1} \sum_l e^{-ikl}
g_l$.

\subsubsection{ndRG generator and scaling equations}
\label{subsec:generator}

Applying the ndRG scheme to the Ising model in Eq.~(\ref{eq:Hamiltonian}), the
disorder amplitude $\sigma$ takes the role of the perturbation strength. At a given UV cutoff $\Lambda$ one obtains for the generator $S\Ls$ of
the unitary transformation, see Eq.~(\ref{eq:generatorDef}):
\be
   S\Ls = \sum_{kq}\LO \left[ \frac{\omega_{kq}}{E_k - E_q} \gamma_k^\dag \gamma_q  + \frac{2m_{kq}}{E_k+E_q} \gamma_k^\dag \gamma_q^\dag- \mathrm{h.c.} \right],
\label{eq:generatorS}
\ee
with the restricted sum defined as
\be
   \sum_{kq}\LO = \sum_{|q|\in[\Lambda,\Lambda+\Delta\Lambda]}\quad \sum_{|k|<\Lambda:|E_k-E_q|>\Omega}.
\ee
Here, we use the notation that capital energies $E_q$ and $E_k$ denote the final renormalized ones after the RG step has been performed. Additionally, we have introduced the scale $\Omega$ that is supposed to
distinguish between resonant $|E_k-E_q|<\Omega$ and off-resonant processes
$|E_k-E_q|>\Omega$, i.e., $\Omega \sim \hbar/t$, according to the energy--time uncertainty relation in Eq.~(\ref{eq:uncertainty}). When physical quantities have been
calculated we replace $\Omega$ by $\Omega =C\hbar/t$  in the end with $C$ a nonuniversal
constant. By keeping $\Omega$ instead of $C\hbar/t$ we are able to identify whether some
properties depend on the nonuniversal details of the RG cutoff and have
therefore to be interpreted with care.

Using Eq.~(\ref{eq:recipeHamRen}) from the ndRG recipe, we obtain at the critical point the following RG equations for the disorder strength $\sigma^2$ and the low-energy spectrum $\varepsilon_k$:
\begin{align}
	\frac{d \sigma^2}{d\Lambda} & = - \frac{\sigma^4}{\pi \Lambda^2}, \nonumber \\
	\frac{d \varepsilon_k}{ d\Lambda} & = \left\{ \begin{array}{ll}
	                                               \frac{2 \sigma^2}{\pi} \frac{\varepsilon_k}{J\Lambda^2}, & \mathrm{ for } \,\, |k|>\sigma^2 \\
	                                               ~ & ~\\
	                                               \frac{2\sigma^2}{\pi} \frac{1}{\Lambda-\varepsilon_k/J} & \mathrm{ for } \,\, |k|<\sigma^2
	                                              \end{array}
	                                              \right.
	\label{eq:finalRGEq}
\end{align}
which we derive in detail below in Sec.~\ref{sec:derivationSpectrum} and Sec.~\ref{sec:derivationDisorder}. Before, however, we aim at discussing shortly their main consequences.

Let us first discuss the renormalization of the spectrum $\varepsilon_k$ for not too small UV cutoffs $\Lambda \gg \sigma^2$. To recapitulate, for $|k|>\sigma^2$, the initial spectrum is approximately linear with $\varepsilon_k \approx J k$ whereas for $|k|<\sigma^2$ we have that $\varepsilon_k \approx \sigma^2/2$ is constant, see Eqs.~(\ref{eq:diagonal_H0}) and (\ref{eq:field_analytical_parametrization}). According to Eq.~(\ref{eq:finalRGEq}), the linear spectrum $\varepsilon_k = v |k|$ of the modes $|k|>\sigma^2$ is modified due to the RG  only such that its velocity $v(\Lambda)=J \exp[-2\sigma^2/(\pi\Lambda)]$ obtains weak perturbative corrections for $\Lambda > \sigma^2$. Here, we have neglected the flow of the disorder strength $\sigma^2$. As we will analyze below, in this regime of the UV cutoff, the disorder strength only acquires weak perturbative corrections which will only contribute beyond second order in the scaling equation for the spectrum.  Similarly, for $|k|<\sigma^2$, the scaling equation~(\ref{eq:finalRGEq}) can also be solved analytically. This yields $\varepsilon_k(\Lambda) = J\Lambda - 2\sigma^2/\pi - 2(\sigma^2/\pi) \mathrm{LW}\{\exp[(\Lambda-\varepsilon_k/J)\pi/(2\sigma^2)]]\} $, with $\mathrm{LW}(x)$ the Lambert-W function and $\varepsilon_k(\Lambda)$ is consequently a monotonously decreasing function of $\Lambda$. Therefore, the initial ndRG flow for the spectrum only leads to perturbative corrections at $\Lambda \gg \sigma^2$ which we will neglect in the following analysis.

When $\Lambda \to \sigma^2/2$ where $\varepsilon_\Lambda \to \varepsilon_k$ for all $|k|<\Lambda$, however, we have to stop the ndRG transformation because the remaining modes $k$ are now all \emph{resonant}. Thus, according to the ndRG recipe given in Sec.~\ref{sec:ndRGRecipe} there are now no off-resonant processes that can be integrated out perturbatively and the UV cutoff $\Lambda_\ast$ where the ndRG comes to an end is given by
\be
	\Lambda_\ast = \frac{\sigma^2}{2}.
\label{eq:Lambda_ast}
\ee
Importantly, however, the \emph{dynamics} of the system is still accessible when stopping the ndRG transformation at this cutoff $\Lambda_\ast$. The reason for that is discussed in Sec.~\ref{sec:ndRGResonantProcesses}.

Notice that although the RG equation for the spectrum does not contain any randomness, this does not mean that the randomness is
fully gone. In fact, it is hidden in the unitary transformation connecting the extended states of the clean with the
localized wavefunctions of the disordered system.

As opposed to the spectrum, the disorder strength increases during the ndRG. According to Eq.~(\ref{eq:finalRGEq}) we find that
\be
	\sigma^2(\Lambda) = \frac{\sigma^2}{1-\frac{\sigma^2}{\pi \Lambda}},
\ee
displaying a strong-coupling divergence for $\Lambda \to \sigma^2/\pi$. Before approaching this strong-coupling divergence the ndRG, however, stops, see Eq.~(\ref{eq:Lambda_ast}). Most importantly, this leaves us within the regime of validity of the current weak-disorder treatment.

In the following, we now aim at showing the derivation of the scaling equations in Eq.~(\ref{eq:finalRGEq}).

\subsubsection{Spectrum}
\label{sec:derivationSpectrum}

We start by determining the renormalized spectrum. Based on Eq.~(\ref{eq:recipeHamRen}) of the ndRG recipe, $S\Ls$ in Eq.~(\ref{eq:generatorS}) generates the following RG equation for the energies:
\begin{align}
	E_k & =\varepsilon_k + \sum_q \LO \left[
	\frac{4|m_{kq}|^2}{E_k+E_q} - \frac{|\omega_{kq}|^2}{E_q-E_k} \right], \nonumber \\
	E_q & =\varepsilon_q + \sum_k \LO \left[
	\frac{4|m_{kq}|^2}{E_k+E_q} + \frac{|\omega_{kq}|^2}{E_q-E_k}\right].
\label{eq:energiesScaling}
\end{align}
As $|E_k-\varepsilon_k| \sim \Delta \Lambda$ (but not $|E_q-\varepsilon_q|$) one can replace $E_k$ by $\varepsilon_k$ on the right-hand side of the above equations. In the following we show how to evaluate the sums appearing in the scaling equations in Eq.~(\ref{eq:energiesScaling}). For illustration we take
\be
	\sum_q \frac{|4m_{kq}|^2}{E_q+\varepsilon_k},
\ee
which is a sum of $|q| \in [\Lambda,\Lambda+\Delta \Lambda]$ over the random variables $m_{kq}$. As is the case in Wilson RG schemes the width of the momentum shell $\Delta \Lambda$ is small but still large enough to host an extensive number of states. Then, in the equation above, we sum over a large number of random variables. Thus, to leading order we can replace the sum over the random variables by the sum over its mean:
\begin{align}
      \sum_q \frac{|4m_{kq}|^2}{E_q+\varepsilon_k} = \sum_q \frac{4\langle|m_{kq}|^2\rangle_\mathrm{dis} }{E_q+\varepsilon_k}+ \mathrm{corr}.
\end{align}
The corrections involve random variables with an associated probability distribution function that shows that their typical magnitude scales as $\sim \sigma^2/\sqrt{N}$. The mean is zero as their sign fluctuates. The variance, however, is finite and scales $\sim \sigma^4/N$. Comparing this to the variance of the amplitudes $w_{kk'}$ and $m_{kk'}$ that are of the order $\sigma^2/N$ it is clear that the corrections above will only contribute to third order and can therefore be neglected within the current accuracy of second order. Using Eq.~(\ref{eq:wmAmplitudes}) we have that $\langle 4|m_{kq}|^2 \rangle_\mathrm{dis} =  \sigma^2 \sin^2(\theta_k-\theta_q)/N$ yielding for the sum:
\begin{align}
      \sum_q \frac{|4m_{kq}|^2}{E_q+\varepsilon_k} = \frac{2\sigma^2}{N}\sum_q \sin^2(\theta_k-\theta_q) \frac{1}{E_q+\varepsilon_k}.
\end{align}
As $|q| = \Lambda$ up to $\Delta \Lambda$ corrections and all remaining contributions are now smooth functions of $q$ we can then evaluate this expression exactly:
\begin{align}
	\sum_q \frac{|4m_{kq}|^2}{E_q+\varepsilon_k} =\frac{2\sigma^2}{\pi}  \frac{1-\cos(2\theta_k)\cos(2\theta_\Lambda)}{E_\Lambda+\varepsilon_k} \Delta \Lambda .
\end{align}
Here, we have used $\theta_{-\Lambda}=-\theta_\Lambda$, and $\sin^2(\theta_k-\theta_\Lambda) + \sin^2(\theta_k+\theta_\Lambda)=1-\cos(2\theta_k)\cos(2\theta_\Lambda)$.
Inserting this result into Eq.~(\ref{eq:energiesScaling}) and using an analogous analysis for the sums over $|w_{kq}|^2$ then yields the scaling equation for the following scaling equation for the slow modes $k$:
\begin{align}
	\frac{d \varepsilon_k}{d\Lambda}  = & \frac{\sigma^2}{\pi} \frac{1+\cos(2\theta_k)\cos(2\theta_\Lambda)}{E\Ls-\varepsilon_k} - \nonumber \\ & -\frac{\sigma^2}{\pi} \frac{1-\cos(2\theta_k)\cos(2\theta_\Lambda)}{E\Ls+\varepsilon_k}.
\label{eq:energiesScalingSmooth1}
\end{align}
Using the result for the Bogoliubov angles in Eq.~(\ref{eq:theta_k}) we see that we have to distinguish two different cases. For $k>\sigma^2$ where $\theta_k = \pi/4-k/4$ for $k>0$ and $\theta_{-k}=\theta_k$, we have that for the low-energy modes $\varepsilon_k \ll E_\Lambda$:
\be
	\frac{d \varepsilon_k}{d\Lambda} = \frac{2\sigma^2}{\pi} \frac{\varepsilon_k}{E_\Lambda^2}, \quad |k|>\sigma^2.
\ee
When $k<\sigma^2$ on the other hand, $\theta_k = k/(2\sigma^2)$ according to Eq.~(\ref{eq:theta_k}) such that:
\be
	\frac{d \varepsilon_k}{d\Lambda} = \frac{2\sigma^2}{\pi} \frac{1}{E_\Lambda-\varepsilon_k}, \quad |k|<\sigma^2.
\ee
When neglecting the weak renormalization of the high-energy modes in the linear regime, i.e. $E\Ls \approx J \Lambda$, then this gives Eq.~(\ref{eq:finalRGEq}).
Additionally, we have to determine the renormalization of the eliminated fast modes whose energies $E_q$ have to be chosen according to the self-consistency equation:
\begin{align}
	E_q =\varepsilon_q & + \frac{\sigma^2}{N} \sum_{k>0} \LO \left[
	\frac{1-\cos(2\theta_k)\cos(2\theta_q)}{E_q+\varepsilon_k}+ \right. \nonumber \\  & \left. + \frac{1+\cos(2\theta_k)\cos(2\theta_q)}{E_q-\varepsilon_k}\right].
\label{eq:energiesScalingSmooth2}
\end{align}
In the following, we will neglect the influence of this renormalization of the eliminated modes $E_q$. First of all, the shift of $E_q$ compared to $\varepsilon_q$ is perturbative in the disorder strength $\sigma$. But most importantly, the eliminated modes are now completely decoupled from the remaining ones and therefore they do not influence the further RG transformation. Being only interested in the leading behavior as a function of $\sigma$, we can therefore neglect this final renormalization and replace $E_q$ by  $\varepsilon_q$ in the following.


\subsubsection{Disorder strength}
\label{sec:derivationDisorder}


Now, we derive the scaling equation, see Eq.~(\ref{eq:finalRGEq}), for the disorder strength. For that purpose, it is suitable to analyze the superpositions
\be
	f_{kk'} = w_{kk'} -2 m_{k -k'}, \quad h_{kk'} = w_{kk'} +2 m_{k -k'},
\ee
rather than the amplitudes $w_{kk'}$ and $m_{kk'}$ themselves. The reason for that is twofold. First, initially, before we start the flow, these two functions can be easily connected to the disorder strength via
\be
	f_{kk'} = g_{k-k'} e^{i\theta_k + i\theta_{k'}}, \quad h_{kk'} = g_{k-k'} e^{-i\theta_k - i\theta_{k'}}.
\ee
These new random variables have zero mean, but a nonzero variance that is solely given by the initial disorder strength:
\be
	\langle |f_{kk'}|^2 \rangle_\mathrm{dis} = \langle |h_{kk'}|^2 \rangle_\mathrm{dis} = \langle |g_{k-k'}|^2|\rangle_\mathrm{dis} =  \frac{\sigma^2}{N}.
	\label{eq:HFinitial}
\ee
Using the generator $S\Ls$ it is straightforward to determine their change under one RG step for low energies when $\varepsilon_k,\varepsilon_{k'} \ll E_\Lambda$:
\begin{align}
	f_{kk'} \mapsto f_{kk'} - \frac{1}{E_\Lambda} \sum_q f_{kq} h_{k'q}^\ast, \nonumber \\ h_{kk'} \mapsto h_{kk'} - \frac{1}{E_\Lambda} \sum_q h_{kq} f_{k'q}^\ast.
\end{align}
As a consequence the variances transform as
\begin{align}
	\langle |f_{kk'}|^2\rangle_\mathrm{dis} \mapsto \langle |f_{kk'}|^2 \rangle_\mathrm{dis} +  \frac{1}{E\Ls^2} \,\,\sum_{q}\,\, \langle |f_{kq} |^2 \rangle_\mathrm{dis} \,\, \langle h_{k'q} |^2 \rangle_\mathrm{dis}, \nonumber \\
	\langle |h_{kk'}|^2\rangle_\mathrm{dis} \mapsto \langle |h_{kk'}|^2 \rangle_\mathrm{dis} +  \frac{1}{E\Ls^2} \,\,\sum_{q}\,\, \langle |h_{kq} |^2 \rangle_\mathrm{dis} \,\, \langle f_{k'q} |^2 \rangle_\mathrm{dis}.
\end{align}
As both the variances of $f_{kk'}$ and $h_{kk'}$ do not have a momentum dependence initially they cannot acquire one during these scaling equations. Thus, one can introduce the functions $F=N\langle |f|^2 \rangle_\mathrm{dis}$ and $H=N\langle |h|^2 \rangle_\mathrm{dis}$ that obey the same scaling equations accordingly:
\be
	\frac{d F}{d\Lambda} = \frac{dH}{d\Lambda}= -\frac{FH}{\pi E\Ls^2}
\ee
As both $H$ and $F$ have the same initial condition, see Eq.~(\ref{eq:HFinitial}), we have that $H=F$ for all $\Lambda$ and it follows that
\be
	\frac{dF}{d\Lambda} = -\frac{F^2}{\pi E\Ls^2}.
\ee
As $F$ characterizes the width of the distribution of the matrix elements $m_{kk'}$ and $w_{kk'}$ we identify $F=H=\sigma^2$ which gives Eq.~(\ref{eq:finalRGEq}).


\subsubsection{Resonant processes}
\label{sec:ndRGResonantProcesses}

In the last sections we have constructed the ndRG transformation and we have discussed the resulting scaling equations for the spectrum and the disorder strength. In order to obtain the dynamics, however, we additionally have to study the influence of the remaining resonant processes. To recapitulate the ndRG recipe in Sec.~\ref{sec:ndRGRecipe}, at the end of the ndRG transformation, the time-evolution operator $P(t)$ can be represented in the form:
\be
	P(t) = U^\dag_\ast e^{-iH\knot_\ast t} e^{-i V\res_\ast t} U.
\ee
In the following we will give a detailed analysis of the resonant processes contained in $V\res_\ast$. First, we outline that $V\res_\ast$ can be diagonalized approximately and therefore its time evolution can be determined analytically which is remarkable because resonant processes typically hamper analytical treatments due to their nonperturbative nature. Secondly and most importantly, it will be shown that this leads to a dynamical classification of the relevance of a perturbation. In particular, it will be shown that for times $t<\sigma^2/(\hbar J)$, the resonant processes are \emph{irrelevant} whereas for times $t>\sigma^2/(\hbar J)$ they become \emph{relevant}.

For that purpose let us analyze the spectrum of the resonant processes $V\res_\ast$. As $V\res_\ast$ only contains resonant contributions we have that
\be
V_\ast\res=\sum_{kk'} \overline{g}_{kk'}c_k^\dag c_{k'}, \,\, \overline{g}_{kk'}=g_{k-k'}\Theta(\Omega-|\varepsilon_k-\varepsilon_{k'}|),
\ee
with $\Theta(x)$ the Heaviside step function. The diagonalization of the full $V_\ast\res$
is difficult. Concentrating onto the low-energy degrees of freedom that are supposed to
contribute dominantly at large times, analytical insight can be obtained. At low energies we have that $\varepsilon_k \approx J k$ for $|k|>\sigma^2$, see Eqs.~\ref{eq:diagonal_H0} and (\ref{eq:field_analytical_parametrization}). For $|k|<\sigma^2$ on the other hand the spectrum $\varepsilon_k= \sigma^2/2$ becomes constant and all modes with $|k|<\sigma^2$ are resonant such they are all contained in $V\res_\ast$. For the moment let us consider for $\overline{g}_{kk'}$ only those modes with $J|k-k'|<\Omega$ for all $k$ and $k$ and neglect the remaining resonant contributions for $|k|<\sigma^2$. The influence of the latter ones will be analyzed below. Using these approximations we have that $\overline{g}_{kk'}\approx g_{k-k'} \Theta(\Omega-J|k-k'|)$. Then, $V_\ast\res$
can be diagonalized analytically via Fourier transformation yielding
\be
	V\res_\ast = \sum_{l} g_{l}(\Omega) c_l^\dag c_l,
	\label{eq:VresastDiagonal}
\ee
with transverse fields $g_l(\Omega)$ depending on the width $\Omega$ of the resonant processes. The probability distribution $P(g,\Omega)$ of the random diagonal elements $g=g_l(\Omega)$ is Gaussian due to the central limit theorem:
\be
   P(g,\Omega) = \sqrt{\frac{\pi}{\sigma^2 J \Omega}} e^{-g^2/(4\sigma^2 J \Omega)}.
\label{eq:pdf}
\ee
This distribution $P(g,\Omega)$ becomes increasingly narrow on large times because $\Omega \sim \hbar/t$ and therefore the typical magnitude $g_\mathrm{typ}\sim \sqrt{\sigma^2 J\Omega} \sim \sqrt{\hbar \sigma^2 J/t}$ of the effective local transverse fields decreases for increasing time. It is important to emphasize, however, that this does \emph{not} imply that these contributions are irrelevant. In particular, it is not the bare fields $g_l(\Omega)$ that determine the time evolution of the resonant contributions but rather the \emph{product} $g_l(\Omega) t$. Consequently, considering this product we find that for times $t>\hbar /J\sigma^2$ the
resonant processes become \emph{relevant} because $g_\mathrm{typ}t/\hbar \sim \sqrt{ \sigma^2 J t/\hbar}$ becomes of order $\mathcal{O}(1)$ when $t=\hbar /J\sigma^2$.

As emphasized before, $V\res_\ast$ in Eq.~(\ref{eq:VresastDiagonal}) does not contain all the resonant processes. In particular, those modes with $|k|<\sigma^2$ have only been partially considered. However, it is the aim of the following discussion to show that these remaining contributions $V_\mathrm{\Lambda_\ast}$ only yield a perturbative correction to $V\res_\ast$ which is beyond the current accuracy of second order in $\sigma$. In terms of the fermionic operator $c_k$, see Eq.~(\ref{eq:c_k}), we have that
\be
	V_\mathrm{\Lambda_\ast}= \sum_{|k|,|k'|<\Lambda_\ast} g_{k-k'} c_k^\dag c_{k'}.
\ee
Because both of the sums over $k$ and $k'$ run over a range proportional to $\Lambda_\ast$ we get that $V_\mathrm{\Lambda_\ast}$ contributes to $\mathcal{O}(\Lambda_\ast^2)$. As we have that $\Lambda_\ast=\sigma^2/2$, see Eq.~(\ref{eq:Lambda_ast}), this gives corrections of the order $\mathcal{O}(\sigma^4)$ to the Hamiltonian in Eq.~(\ref{eq:VresastDiagonal}) which can be neglected within the current accuracy.





\subsection{Results}
\label{sec:results}

The main results have already been summarized in Sec.~\ref{sec:resultsSummary}. In the following, we outline their derivation using the ndRG and discuss their physical implications in detail. First, we consider the localization properties in Fock space by analyzing the dynamics of the Loschmidt echo in Sec.~\ref{sec:FockSpace}. Afterwards, we study localization in real space via the local memory in Sec.~\ref{sec:realSpace}.

\subsubsection{Localization in Fock space: Loschmidt echo}
\label{sec:FockSpace}

In order to evaluate the Loschmidt echo $\mathcal{L}(t)$, defined in Eq.~(\ref{eq:defLoschmidtEcho}), we use that the ndRG provides the following representation of the time-evolution operator:
\be
   P(t) = U_\ast^\dag e^{- i H_\ast t/\hbar} U_\ast, H_\ast = H\knot_\ast + V\res_\ast,
\ee
Commuting the exponential $\exp[-i H_\ast t/\hbar]$ past the unitary transformation $U_\ast^\dag$ we find that the Loschmidt echo can be written as:
\be
     \mathcal{L}(t) = \left|\left\langle \psi_0 \left| U_\ast^\dag(t) U_\ast \right| \psi_0 \right\rangle \right|^2,
\ee
with $U_\ast^\dag(t) = \exp[iH_\ast t] U_\ast^\dag \exp[-i H_\ast t]$. In order to arrive at this identity we have used that time evolution with the Hamiltonian $H_\ast$ leaves the initial state $|\psi_0\rangle$ invariant up to a phase:
\be
	e^{- i H_\ast t} |\psi_0\rangle \stackrel{t \to \infty}{\longrightarrow} e^{-i E_\ast t} | \psi_0\rangle,
\ee
with $E_\ast\in \mathbb{R}$. This acquired phase, however, does not contribute to the Loschmidt echo due to the modulus taken. In the following, we show the derivation of this property. First of all, due to the factorization property of the resonant processes, see Eq.~(\ref{eq:recipeTEOF}), we have that:
\be
	e^{-i H_\ast t/\hbar} = e^{-i H\knot_\ast t/\hbar} e^{-i V\res_\ast t/\hbar},
\ee
such that we can address the time evolution with $H\knot_\ast$ and $V\res_\ast$ separately. The initial state $|\psi_0\rangle$ is, by construction of the ndRG, an eigenstate of $H\knot_\ast$ such that $\exp[-iH\knot_\ast t]| \psi_0 \rangle =\exp[-iE_\ast t] |\psi_0\rangle$. But even more importantly, it is necessary to estimate the influence of the resonant processes:
\be
	e^{iV\res_\ast t} |\psi_0\rangle,
\ee
with
\be
	V\res_\ast = \sum_{kk'}\!^{\!^\Omega} \left[ w_{kk'} \gamma_k^\dag \gamma_{k'} + \left( m_{kk'} \gamma_k^\dag \gamma_{k'}^\dag + h.c.\right) \right].
\ee
Here, $\sum_{kk'}\!\!\!\!\!\!^{\!^\Omega}$ denotes the sum over all momenta $k$ and $k'$ such that $|E_k-E_{k'}|<\Omega$. Importantly, $m_{kk'}\propto |k-k'|$ for $|k-k'|\to 0$, see Eq.~(\ref{eq:wmAmplitudes}), in contrast to $w_{kk'}$ such that asymptotically $V\res_\ast \to \sum_{kk'}\!\!\!\!\!\!^{\!^\Omega} \,\, w_{kk'} \gamma_k^\dag \gamma_{k'}$ for $t\to \infty$. As the initial state $|\psi_0\rangle$ is the vacuum for the $\gamma_k$ operators by construction, i.e., $\gamma_k |\psi_0\rangle=0$, we have that $V\res_\ast |\psi_0\rangle \to 0$ in the long-time limit. Therefore $\exp[-i V\res_\ast t] |\psi_0\rangle = |\psi_0\rangle$. Notice that the property, that $V\res_\ast$ does not induce time evolution of the vacuum $|\psi_0\rangle$, does not imply that its overall dynamics can be neglected. In particular, it will be shown below that the excitations on top of $|\psi_0\rangle$ contained in the unitary transformation $U_\ast$ are strongly affected by it.


As $U_\ast = \mathcal{T}\Ls \exp[\int d\Lambda S\Ls]$, see Eq.~(\ref{eq:Uast}), and $S\Ls$ by construction only contains perturbative processes, one can perform a cumulant expansion~\cite{Kubo1962xh} up to second order in the disorder strength $\sigma$ yielding:
\be
   \mathcal{L}(t) = \left| \exp\left[ \int d\Lambda d\Lambda' \langle (S\Ls-S\Ls(t) ) S_{\scriptscriptstyle{\Lambda'}} \rangle \right] \right|^2,
\ee
with $\langle \dots \rangle = \langle \psi_0|\dots|\psi_0 \rangle$. For the final evaluation of this expression it is necessary to determine the dynamics under the resonant contributions $V_\ast\res$ for which all the necessary steps have been presented already in Sec.~\ref{sec:ndRGResonantProcesses}. In terms of the Jordan-Wigner fermions $c_k$, that are connected to the quasiparticles $\gamma_k$ via $c_k=\cos(\theta_k)\gamma_k-i \sin(\theta_k)\gamma_{-k}^\dag$, we have that $V_\ast\res \approx \sum_{kk'} g_{k-k'} \Theta(\Omega-J|k-k'|) c_k^\dag c_{k'}$. This can be diagonalized analytically by Fourier transformation yielding $V_\ast\res = J \sum_{l} g_l(\Omega) c_l^\dag c_l$ where the $g_l(\Omega)$ are random variables with an associated probability distribution $P(g)=\sqrt{\frac{\pi J}{\sigma^2 \Omega}} \exp[-g^2 J/(4\sigma^2 \Omega)]$, see Eq.~(\ref{eq:pdf}). Thus, we have that $e^{iV_\ast\res t}c_ke^{-iV_\ast \res t} = \sum_{k'}\alpha_{kk'}(t) c_{k'}$ with $\alpha_{kk'}(t) = N^{-1} \sum_l e^{-i(k-k')l}e^{-ig_l(\Omega) t }$. For the diagonal element $\alpha_{kk}(t)$ one obtains then, for example, after disorder averaging $\langle \alpha_{kk}(t)\rangle_\mathrm{dis} = e^{-\sigma^2J\Omega t^2/\hbar^2}$. Using Eq.~(\ref{eq:generatorS}) for $S\Ls$ this gives for the Loschmidt echo rate function $\lambda(t)$ after disorder averaging:
\begin{align}
   \overline{\lambda}(t) = \langle \lambda(t) \rangle_\mathrm{dis} = & -\frac{\sigma^2}{\pi^2}\int_{\Lambda_\ast}^{\pi} dk \int_0^k dq \, T_{kq} \times \nonumber \\ & \times \frac{\cos[(\varepsilon_k+\varepsilon_q)t/\hbar]a^2(t)-1}{\left(\varepsilon_q+\varepsilon_k \right)^2},
\label{eq:Loschmidt}
\end{align}
for times $t\gg \hbar/J\sigma$ with $T_{kq}=1-\cos(2\theta_k)\cos(2\theta_q)$ and $\Lambda_\ast$ the final value of the UV cutoff. The influence of the resonant contributions is contained in the function $a(t)$ that has the property $a(t)\to 1$ for $\hbar/J\sigma \ll t \ll \hbar /J\sigma^2$ and $a(t)\to 0$ for $t\gg \hbar /J\sigma^2$. Using Eq.~(\ref{eq:pdf}) the functional form of $a(t)$ can be estimated as $a(t)=\exp[-\sigma^2 J \Omega t^2/\hbar^2]$ decaying exponentially as a function of time because $\Omega \sim \hbar/t$. As the dynamics during the crossover at a time scale $t\sim \hbar/J\sigma^2$ depends on the details of the RG cutoff function via $\Omega$ we expect that this crossover cannot be described quantitatively.

\begin{figure}
\centering
\includegraphics[width=\linewidth]{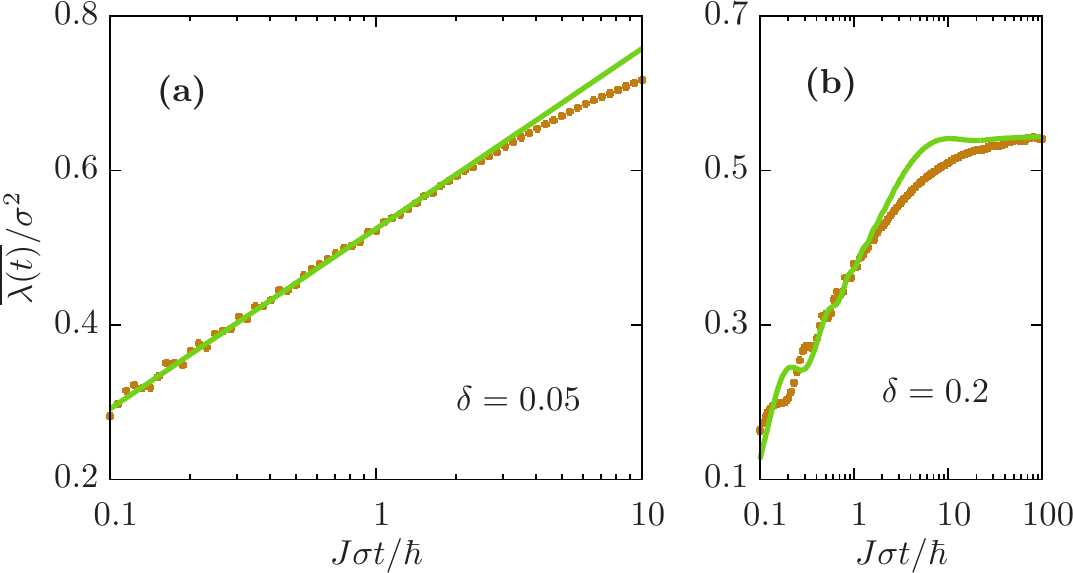}
\caption{(color online) Dynamics of the Loschmidt rate function
$\overline{\lambda(t)}$. (a) Logarithmic growth of $\overline{\lambda}(t)$ on intermediate time scales. Comparison of the numerically exact solution (points) to the
logarithmic growth $(\sigma/\pi)^2\log(2JtC/\hbar)$, $C=\mathrm{const.}$, found by the
analytics (line) on intermediate times for $\delta=0.05$. For the exact numerics data is shown for  $N=3200$ averaged over $1500$ realizations of field disorder. The constant $C$, only a subleading contribution for $\sigma\to0$, has been obtained by a fit to the numerical data. (b) Comparison of the
full analytical result of Eq.~(\ref{eq:Loschmidt})  with the exact
numerics on all time scales. Here, we have used $\delta=0.2$, $N=3200$, and $1500$ realizations of disorder. The agreement is very good except during the
crossover at times $t\approx\hbar/J\sigma^2$ where the analytical result explicitly depends on the RG cutoff $\Omega$. }
\label{fig:Loschmidt}
\end{figure}

Fig.~\ref{fig:Loschmidt} shows a comparison of the result in Eq.~(\ref{eq:Loschmidt}) with the exact numerics obtained by the extensive simulations for large systems up to $N=3200$ using the  methods outlined in Ref.~\onlinecite{Jacobson2009}. As one can see, the analytical result nicely matches the exact solution except at the crossover time scale $t\approx \hbar/J\sigma^2$ where the analytical result depends explicitly on the RG cutoff $\Omega$ .

Using the result in Eq.~(\ref{eq:Loschmidt}), we find that on intermediate times $\hbar/J\sigma \ll t \ll \hbar /J\sigma^2$ the Loschmidt echo rate function
$\overline{\lambda(t)}$ shows a slow logarithmic growth
\be
      \overline{\lambda}(t) = (\sigma/\pi)^2\log(2Jt/\hbar),
\ee
as already mentioned before in Sec.~\ref{sec:resultsSummary}. This analytical result shows very good agreement with the exact numerics, see Fig.~\ref{fig:Loschmidt}. An increasing Loschmidt echo rate function characterizes an increasing deviation of the time-evolved state from its initial Fock state. We attribute the particularly slow growth of $\overline{\lambda}(t)$ to the expected nonergodicity and Fock-space localization at the infinite-randomness fixed point~\cite{Vosk2013}.

For even longer times $t \gg \hbar /J\sigma^2$, the resonant processes become of particular importance. As mentioned already before, they are responsible for the decay of the function $a(t) \to 0$ for times $t\gg \hbar/J\sigma^2$. From Eq.~(\ref{eq:Loschmidt}) we find that as a consequence the Loschmidt rate function $\overline{\lambda}(t)$ saturates to a constant value given by:
\be
	\lambda_\infty=\overline{\lambda}(t\to\infty) = (\sigma/\pi)^2 \log(\sigma^2).
	\label{eq:lambda_infty}
\ee
Remarkably, this result is nonperturbative in the disorder strength, displaying a divergent second derivative in the limit of vanishing disorder. This nonanalytic behavior of $\lambda_\infty$ serves as an indicator for the instability of the quantum phase transition in the transverse-field Ising chain against disorder. This is a consequence of the Harris criterion~\cite{Harris1974} when applied to the current model system. Although the switched-on disorder in our quantum quench scenario might be arbitrarily weak for $\sigma \to 0$, the low-energy degrees of freedom in the disordered Ising chain which are probed for asymptotically large times are not adiabatically connected to those of the homogeneous model.
\begin{figure}
\centering
\includegraphics[width=\linewidth]{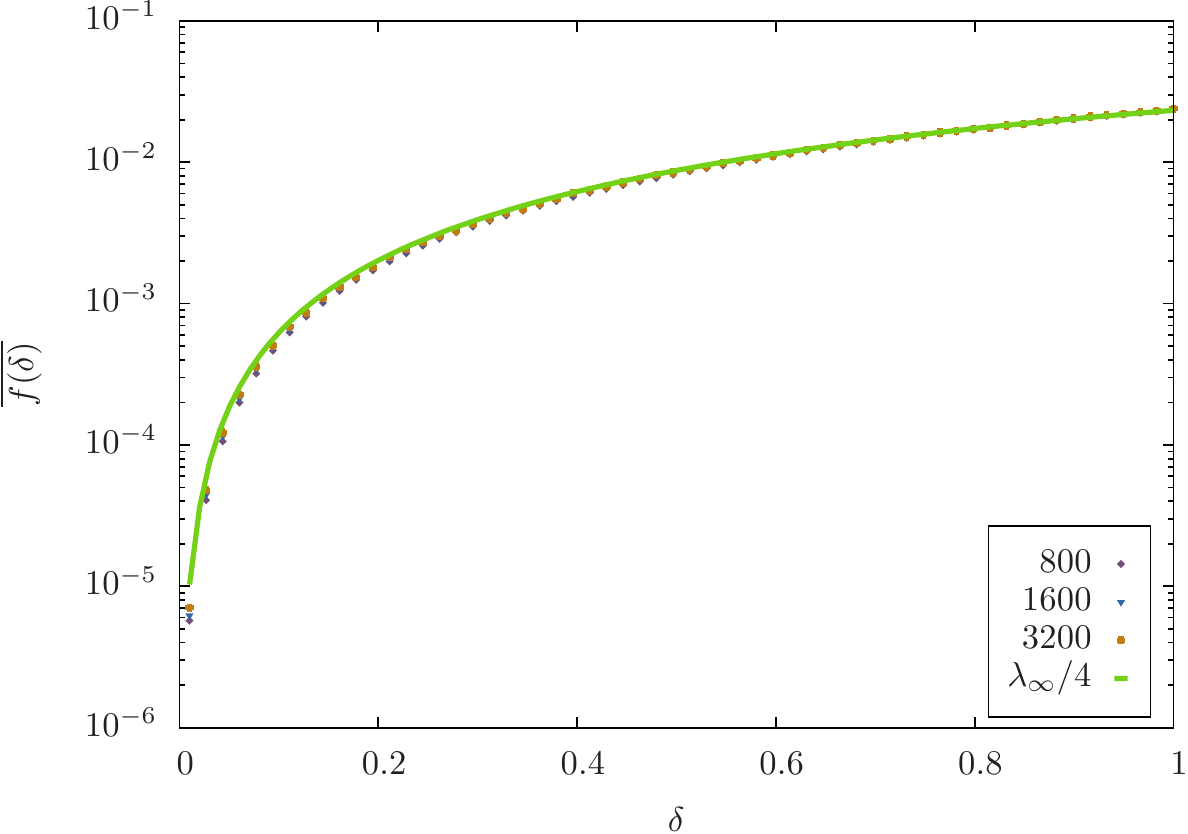}
\caption{(color online) Relating the asymptotic long-time value $\lambda_\infty = \overline{\lambda}(t\to\infty)$ of the Loschmidt echo rate function to the fidelity rate function $\overline{f}(\delta)$ as a function of the disorder strength $\delta$. Numerical data for $\overline{f}$ (points) for different system sizes up to $N=3200$ and $10^4$ disorder realizations compared to the analytical formula (line) $\lambda_\infty = \sigma^2\log(\sigma^2D)/\pi^2$ with $D$ a
fit parameter and $\sigma^2 =\delta^2/3$. }
\label{fig:fid}
\end{figure}

Moreover, we find that the asymptotic long-time value $\lambda_\infty$ is, remarkably, connected to a static equilibrium property, as we will discuss in the following. Specifically, we also calculated numerically the fidelity $F = |\langle \phi_0 | \psi_0\rangle|$ for our model with $|\phi_0\rangle$ the ground state of the final disordered transverse-field Ising chain. We have that $F=\exp[-Nf]$, in the same way as the Loschmidt echo, shows large deviation scaling. Using the methods outlined in Ref.~\onlinecite{Jacobson2009}, we have determined numerically exactly for large systems the fidelity rate function $f=-N^{-1} \log(F)$. We find that
\be
	\lambda_\infty = 4 \overline{f} = 4 \langle f \rangle_\mathrm{dis},
	\label{eq:lambda_4f}
\ee
by comparing the numerical results for $f$ to the analytical result for $\lambda_\infty$ in Eq.~(\ref{eq:lambda_infty}), see Fig.~\ref{fig:fid}. As one can see, the agreement is very good. Remarkably, this identity relates the Loschmidt echo, that, in principle, experiences the full many-body spectrum, to mere ground-state properties contained in the fidelity which we interpret as an indicator for Fock-space localization and therefore nonergodicity of the final Hamiltonian. A similar observation about the connection between the asymptotic long-time limit of the Loschmidt echo and the Fidelity has been made recently for quantum quenches in integrable and nonergodic Luttinger liquids~\cite{Dora2014}.

An additional interesting implication of the result in Eq.~(\ref{eq:lambda_4f}) is that using Eq.~(\ref{eq:lambda_infty}), the associated fidelity susceptibility~\cite{Venuti07,Gu2009}
\be
	\chi_s = - \left.\frac{d}{d\sigma^2}\right|_{\sigma=0} F(\sigma^2) \,\,\, \stackrel{\sigma \to 0}{\longrightarrow} \,\,\, \frac{1}{\pi} \log(\sigma)
\ee
shows a logarithmic divergence. The fidelity susceptibility characterizes the sensitivity of the ground-state wavefunction against an infinitesimal change of a parameter which is $\sigma$ in the present case. As a consequence, disorder, although arbitrarily weak, leads to a drastic change of the ground-state wavefunction at the quantum critical point  of the homogeneous Ising chain.  This is in perfect agreement with the Harris criterion according to which this quantum critical point is unstable against disorder as already discussed below Eq.~(\ref{eq:lambda_4f}).

Remarkably, the ndRG is capable to obtain this nonanalytic behavior in the perturbation strength. This is possible because the resonant processes can be accounted for explicitely,  which distinguishes the ndRG from other RG approaches.


\subsubsection{Localization in real space: local memory}
\label{sec:realSpace}

After having discussed Fock-space localization properties, we now turn to a study of the localization dynamics in real space. As already outlined in Sec.~\ref{sec:resultsSummary} we are interested in the dynamics of the local memory which can be characterized via the autocorrelation function
\be
   \chi(t) = \left\langle \frac{1}{N} \sum_{l=1}^N \langle \sigma_l^x(t) \sigma_l^x \rangle_c \right\rangle_\mathrm{dis}
\ee
where $\langle  \sigma_l^x(t) \sigma_l^x  \rangle_c = \langle \sigma_l^x(t) \sigma_l^x  \rangle-\langle  \sigma_l^x(t) \rangle\langle \sigma_l^x  \rangle$ denotes the cumulant and $\langle \dots \rangle=\langle \psi_0|\dots |\psi_0\rangle$.

The main results obtained for $\chi(t)$ have already been summarized in Sec.~\ref{sec:resultsSummary}. Here, we will show their derivation and we will discuss the results in detail. In particular, we will be interested on intermediate time scales $\hbar/\sigma J < t < \hbar/\sigma^2 J$. Using the Jordan-Wigner transformation and a subsequent Fourier transformation the autocorrelator $\chi(t)$ can be written in the following way:
\begin{align}
   \chi(t) =  \frac{1}{N^2} \sum_{kk'q} \left[  \left\langle \langle c_{k+q}^\dag(t) c_k(t) c_{k'}^\dag c_{k'+q} \rangle \right\rangle_\mathrm{dis}
   - \right. \nonumber \\ \left. - \left\langle \langle c_{k+q}^\dag(t) c_k(t)\rangle \langle c_{k'}^\dag c_{k'+q}  \rangle \right\rangle_\mathrm{dis} \right].
\end{align}
In order to obtain the dynamics of the fermionic operators $c_k$ we have to determine the action of the unitary transformation $U_\ast$ onto the quasiparticles $\gamma_k$ which are connected to the $c_k$ operators via a Bogoliubov rotation, see Eq.~(\ref{eq:c_k}). To lowest order in $\sigma$ we have that:
\be
   U^\dag_\ast \gamma_k U_\ast = \gamma_k + \sum_{|q|=|k|}^{\pi} \!\!\!\!\! ~^{^\Omega} \left[\frac{w_{kq}}{\varepsilon_k - \varepsilon_q} \gamma_q + \frac{2m_{kq}}{\varepsilon_k+\varepsilon_q} \gamma_q^\dag\right].
   \label{eq:U_gamma_k}
\ee
As before, the superscript $\Omega$ in $\sum_{|q|=|k|}^{\pi} \!\!\!\!\!\!\!\!\!\!\!\!\!\! ~^{^\Omega}\,\,\,\,\,\, $ is supposed to mean a sum over $q$ such that $|\varepsilon_q-\varepsilon_k|>\Omega$. Based on this result, we can decompose the autocorrelator into:
\be
   \chi(t) = \chi\knot(t) + \chi_d(t)
\ee
where $\chi\knot(t)$ only contains the zeroth-order contribution of the transformed $\gamma_k$ operators in Eq.~(\ref{eq:U_gamma_k}), i.e., all those without the $w_{kq}$ and $m_{kq}$ terms. The remaining contributions are collected in $\chi_d(t)$ accordingly. In the following, we will analyze $\chi\knot(t)$ and $\chi_d(t)$ separately.

\begin{figure}
\centering
\includegraphics[width=\linewidth]{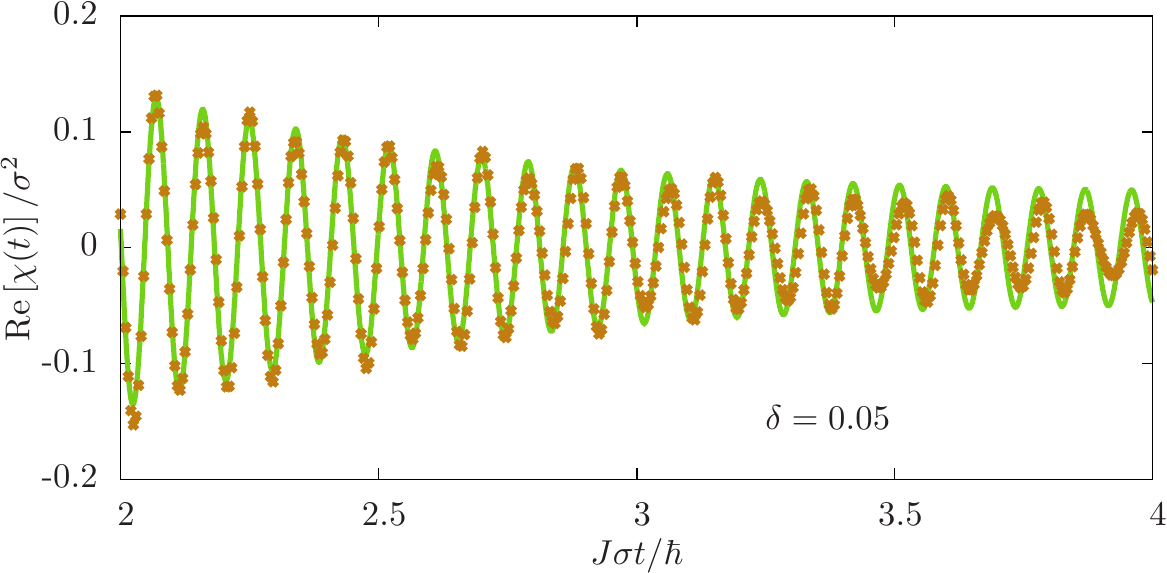}
\caption{(color online) Autocorrelation function $\chi(t)$ on intermediate time scales $\hbar / (J\sigma) < t < \hbar/(J\sigma^2)$. Comparison of exact numerics (points) for $N=2400$ and $1000$ realizations with the analytic asymptotics, see Eq.~(\ref{eq:results_chi_infty}) (line) showing very good agreement. Notice that there is no fit parameter involved. }
\label{fig:chi1}
\end{figure}

\begin{figure}
\centering
\includegraphics[width=\linewidth]{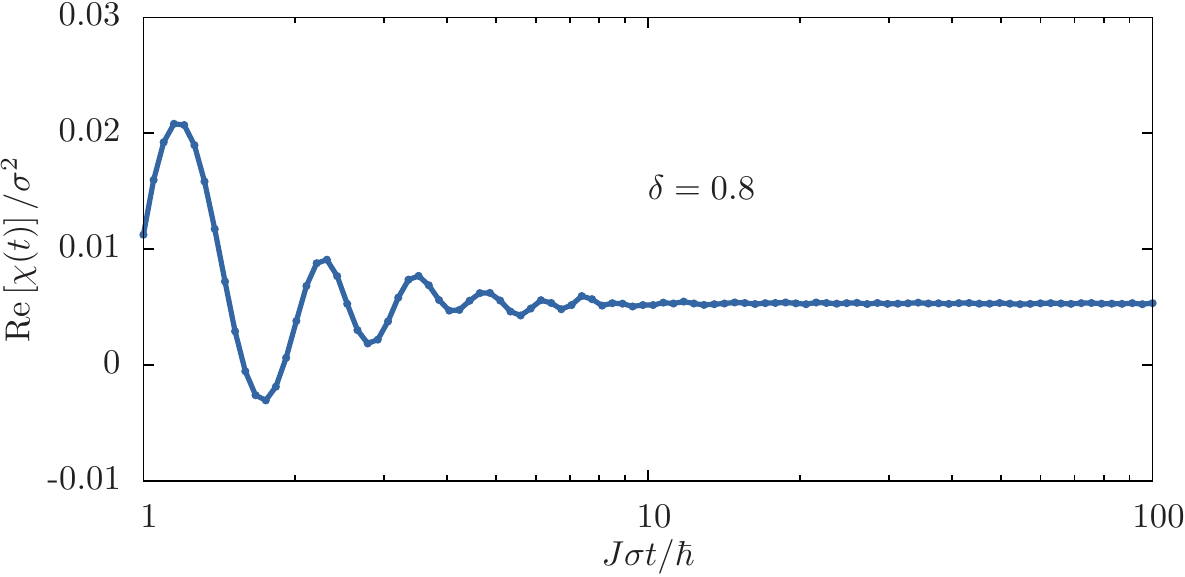}
\caption{(color online) Long-time dynamics of the local memory $\chi(t)$ for $\delta=0.8$, $N=1000$, and $1000$ realizations. For times $t\gg \hbar/(J\sigma^2)$, $\chi(t)$ relaxes to a constant $\chi_\infty = \lim_{t\to\infty} \chi(t)>0$  implying localization and nonergodicity in real space.}
\label{fig:chi2}
\end{figure}

Let us first concentrate on $\chi\knot(t)$. Using Eq.~(\ref{eq:c_k}) for the connection between the $c_k$ and $\gamma_k$ operators one therefore directly obtains that $\chi\knot(t)$ is of product form:
\be
   \chi\knot(t) = \chi\knot_1(t) \chi\knot_2(t),
\ee
with
\begin{align}
    \chi\knot_1(t) & = \int _0^\pi \frac{dk}{\pi} \cos^2(\theta_k) e^{-i\varepsilon_k t} , \nonumber \\
    \chi\knot_2(t) & = \int _0^\pi \frac{dk}{\pi} \sin^2(\theta_k) e^{-i\varepsilon_k t}.
\end{align}
Their long-time asymptotics can be obtained straightforwardly. Using the formulas for the Bogoliubov angles in Eq.~(\ref{eq:theta_k}) this yields:
\begin{align}
	\chi\knot_1(t) & \stackrel{t\to\infty}{\longrightarrow} \frac{1}{\sqrt{\pi t}} e^{-2iJt/\hbar + i\pi/4}, \nonumber \\
	\chi\knot_1(t) & \stackrel{t\to\infty}{\longrightarrow} -\frac{i}{2\pi t}.
\end{align}
As a consequence, this gives for the full $\chi\knot(t)$ the following long-time asymptotics:
\be
	\chi\knot(t) \stackrel{t\to\infty}{\longrightarrow} \frac{1}{2} \frac{e^{-2iJt/\hbar - i\pi/4}}{\left( \pi t\right)^{3/2}}.
	\label{eq:result_chi_knot}
\ee
Having established the dynamics of $\chi\knot(t)$ we now aim at calculating the asymptotics of the remaining contribution $\chi_d(t)$. Using Eq.~(\ref{eq:U_gamma_k}) $\chi_d(t)$ can be written after straightforward algebra in the following form:
\be
	\chi_d(t) = i \frac{1}{N^2} \sum_{k p p'} \sin(\theta_{p'}-\theta_p) \nu_p^{k+p} \nu_{p'}^{k-p'} \left\langle A_{pp'}^k \right\rangle_\mathrm{dis},
\ee
with $\nu_p^k = 0 $ for $|p|<|k|$ and $\nu_p^k=1$ otherwise. The function $A_{pp'}^k$ is defined as
\begin{align}
	 A_{pp'}^k & = \left[ \cos(\theta_{k+p}) B_{k}^p - i \sin(\theta_{k+p}) C_{-k}^p \right] \nonumber \\
	 & \times \left[ \cos(\theta_{k-p'}) C_{k}^{p'} + i \sin(\theta_{k-p'}) B_{-k}^{p'} \right],
	 \label{eq:Appk}
\end{align}
where
\begin{align}
	B_k^p & = w_{k+p,p}\,\, e^{-i \varepsilon_p t}\,\, \frac{1-e^{-i[\varepsilon_{k+p}-\varepsilon_p]t/\hbar}}{\varepsilon_{k+p}-\varepsilon_p} \nonumber \\
	C_k^p & = 2 m_{k-p,p}^\ast \,\, e^{-i\varepsilon_p t} \frac{1-e^{i [\varepsilon_{k-p}+\varepsilon_p]t/\hbar}}{\varepsilon_{k-p}+\varepsilon_p}.
\end{align}
Analyzing all of the contributions in Eq.~(\ref{eq:Appk}) the asymptotic long-time regime is dominated by a single one:
\begin{align}
	\chi_d(t) & \stackrel{t \gg \hbar/J\sigma}{\longrightarrow}  i \frac{1}{N^2} \sum_{k p p'} \sin(\theta_{p'}-\theta_p) \nu_p^{k+p} \nu_{p'}^{k-p'} \nonumber \\ & \times \cos(\theta_{k+p})  \cos(\theta_{k-p'}) \left\langle B_{k}^p C_{k}^{p'} \right\rangle_\mathrm{dis}.
\end{align}
First of all we note that the disorder average can be performed at this point analytically where we use that $\langle w_{k+p,p} 2 m_{k-p',p'}^\ast \rangle_\mathrm{dis} = -i \sigma^2 N^{-1} \cos(2\theta_p) \sin(2\theta_{p'})$, see Eq.~(\ref{eq:wmAmplitudes}). Then it is suitable to analyze the functional dependence of the Bogoliubov angles which allows to isolate the dominant contributions for the long-time asymptotics. The leading behavior of the sum over $p$ one obtains by use of a stationary phase approximation in the vicinity of $p\approx \pi$ while the sum over $p'$ is dominated by the long-wavelength limit $p' \approx 0$. Expanding all appearing functions around $p\approx \pi$ and $p' \approx 0$ one obtains after turning the sums into integrals:
\begin{align}
	\chi_d(t) = 2\sigma^2 I_1(t) I_2(t) I_3(t)
\end{align}
with
\begin{align}
	I_1(t) & = \int \frac{dp}{2\pi} e^{-2iJt/\hbar + i(p-\pi)^2 Jt/4\hbar} \rightarrow \frac{e^{-2iJt/\hbar +i \pi/4}}{\sqrt{\pi J t/\hbar}} ,\nonumber \\
	I_2(t) & = \int \frac{dp}{2\pi}e^{ip Jt/\hbar} \frac{e^{-2ip Jt/\hbar}-1}{2p} \rightarrow -\frac{i}{4}, \nonumber \\
	I_3(t) & = 2\int \frac{dk}{\pi} \frac{e^{-ik^2 Jt/4\hbar}-1}{k^2} \rightarrow -i \sqrt{\frac{Jt}{\hbar \pi}} e^{-i\pi/4}.
\end{align}
Therefore, this yields
\be
	\chi_d(t) \rightarrow -\frac{\sigma^2}{4 \pi} e^{-2iJt/\hbar},
	\label{eq:result_chi_d}
\ee
which is a nondecaying constant (up to an oscillating phase factor). Combining the results for $\chi\knot(t)$ in Eq.~(\ref{eq:result_chi_knot}) and for $\chi_d(t)$ in Eq.~(\ref{eq:result_chi_d}) we find that the full autocorrelator $\chi(t)$ experiences the following decay on time scales $\hbar/J\sigma \ll t \ll \hbar/J\sigma^2$:
\be
	\chi(t) \rightarrow \frac{1}{2} e^{-2iJt/\hbar} \left[ \frac{e^{-i\pi/4}}{(\pi J t/\hbar)^{3/2}} - \frac{\sigma^2}{2\pi} \right],
	\label{eq:results_chi_infty}
\ee
as already presented in Sec.~\ref{sec:resultsSummary}. A comparison of the numerical and analytical result is shown in Fig.~\ref{fig:chi1}, with the agreement being remarkably good. Notice that there is no fit parameter involved. The influence of disorder is solely contained in the static contribution. The algebraic decay $\propto t^{-3/2}$ is also present in the homogeneous system without disorder and originates from the dynamics of effectively freely propagating quasiparticles. For longer times, see Fig.~\ref{fig:chi2} which will be discussed in more detail below, these algebraically decaying oscillations die out, however.

That a non-decaying contribution to $\chi(t)$ in the long-time limit is related to localization and absence of ergodicity can be seen in the following way. In the asymptotic long-time regime, the final state of an ergodic thermalizing system is describable by a canonical ensemble whose only dependence on the initial state is the defining temperature.\cite{Polkovnikov2011kx} Therefore, any initial local information is lost which can be formally expressed by the factorization property $\lim_{t\to \infty}\langle \sigma_l^x(t) \sigma_l^x  \rangle =\lim_{t\to\infty} \langle \sigma_l^x(t) \rangle \langle \sigma_l^x  \rangle$. As we are looking at the connected correlation function this is equivalent to $\chi_\infty =0$. Concluding, a nonzero $\chi_\infty\not=0$ implies localization and absence of ergodicity.

On time scales up to $t \to \hbar/(J\sigma^2)$, the analytical result in Eq.~(\ref{eq:result_chi}) predicts a nondecaying static contribution and therefore localization and nonergodicity. What happens on even longer time scales $t>\hbar/(J\sigma^2)$ is, concerning $\chi(t)$ instead of $\lambda(t)$, beyond the scope of the present second order treatment of the used ndRG. In order to clarify the asymptotic long-time behavior of $\chi(t)$ we have therefore studied this problem numerically, see Fig.~\ref{fig:chi2}. We find that
\be
	\chi_\infty >0,
\ee
again confirming the localized nature of the studied system. The observed localization dynamics in real space reflects the localized nature of final disordered transverse-field Ising chain where the parameters are chosen such that the system is located right at its infinite-randomness critical point. For the data of the numerical simulations in Fig.~\ref{fig:fid}, we have used a slightly larger disorder strength $\delta = 0.8$ in order to be able to reach the asymptotic long-time limit. However, this is still within the regime of weak disorder, as one can see from Fig.~\ref{fig:fid} where the analytical weak-disorder result of the Loschmidt echo rate function obtained using the ndRG is compared to exact numerics. There, the weak-disorder results at a disorder strength of $\delta=0.8$ still fit well.

Summarizing, in this section we have studied localization dynamics and ergodicity in real space on the basis of the local memory $\chi(t)$.
On intermediate times the local autocorrelation function develops a static contribution, see Eq.~(\ref{eq:results_chi_infty}), which is a precursor to localization in the long-time limit. There, we find that $\chi_\infty \not= 0$ implying a nonvanishing local memory. Thus, the system is nonergodic.

\section{Conclusion and Outlook}

In this work we have formulated the nonequilibrium dynamical renormalization group (ndRG) for the analytical description of the nonequilibrium dynamics in quantum many-body systems. Contrary to conventional RG schemes, the ndRG accounts for resonant processes which is important for the description of the long-time dynamics.

We have demonstrated the capabilities of the ndRG by applying it to quantum quenches in a complex and paradigmatic model system, the disordered transverse-field Ising chain. For quantum quenches from the homogeneous to the infinite-randomness critical point we studied the localization dynamics in real as well as many-body Fock space.

In principle, the ndRG can be applied to any weakly perturbed exactly solvable system. Because the ndRG is capable to account for resonant processes, although nonperturbative in nature, it is especially suited to address the long-time dynamics of interacting quantum many-body systems. This encompasses questions of fundamental importance such as thermalization as well as quantum ergodicity~\cite{Polkovnikov2011kx} an therefore also for many-body localization~\cite{Altshuler1997hx,Basko2006,Nandkishore2014,Altman2014}. In this context, it is particularly noteworthy that the ndRG has already been successfully applied for such systems~\cite{Hauke2014mb}.

Moreover, it is important to emphasize that the ndRG is not only applicable to systems subject to a sudden switch of their parameters in terms of a quantum quench but also to other temporal dependencies of the Hamiltonian. In this context, it might be of particular interest to apply the ndRG to periodically driven systems where a novel class of nonequilibrium phase transitions in interacting quantum many-body systems has been discovered recently which have been termed energy-localization transitions~\cite{DAlessio2012}.



\begin{acknowledgments}
The authors thank A. Polkovnikov and S. Kehrein for valuable discussions. This work has
been supported by the DFG (SFB 1143 and GRK 1621), by the Austrian Science Fund FWF (SFB FOQUS
F4016), and by the Deutsche Akademie der Naturforscher Leopoldina under grant number LPDS
2013-07.
\end{acknowledgments}


\bibliographystyle{apsrev}
\bibliography{Heyl_Vojta_ndRG}

\end{document}